\documentclass[aps,prl,superscriptaddress,showpacs,twocolumn]{revtex4-1}
\usepackage[left=1.8cm,right=1.8cm,bottom=1.9cm,top=1.8cm,ignoreall]{geometry}
\usepackage[utf8]{inputenc}
\usepackage{graphicx}
\usepackage{amsmath}
\usepackage{amssymb}
\usepackage{xspace}
\usepackage{bm}
\usepackage{color}
\usepackage[squaren,Gray]{SIunits}

\newcommand*{\Scale}[2][4]{\scalebox{#1}{$#2$}}%
\newcommand{\FB}{\color{black}}
\newcommand{\HPR}{H\!P\!R}

\begin{document}

\title{\FB Bosonic Condensation and Disorder-Induced Localization in a Flat Band}

\author{F. Baboux}
\thanks{florent.baboux@lpn.cnrs.fr}
\affiliation{Laboratoire de Photonique et de Nanostructures (LPN), CNRS, Université Paris-Saclay, route de Nozay, F-91460 Marcoussis, France}

\author{L. Ge}
\affiliation{Department of Engineering Science and Physics, College of Staten Island, CUNY, New York 10314, USA}
\affiliation{The Graduate Center, CUNY, New York 10016, USA}

\author{T. Jacqmin}
\altaffiliation{Present adress: Laboratoire Kastler Brossel, UPMC-Sorbonne Universités, CNRS, ENS-PSL Research University, Collège de France, 4 place Jussieu, Case 74, F75252 Paris Cedex 05, France}

\affiliation{Laboratoire de Photonique et de Nanostructures (LPN), CNRS, Université Paris-Saclay, route de Nozay, F-91460 Marcoussis, France}

\author{M. Biondi}
\affiliation{Institute for Theoretical Physics, ETH Zurich, 8093 Zurich, Switzerland}

\author{E. Galopin}
\affiliation{Laboratoire de Photonique et de Nanostructures (LPN), CNRS, Université Paris-Saclay, route de Nozay, F-91460 Marcoussis, France}

\author{A. Lema\^itre}
\affiliation{Laboratoire de Photonique et de Nanostructures (LPN), CNRS, Université Paris-Saclay, route de Nozay, F-91460 Marcoussis, France}

\author{L.~Le~Gratiet}
\affiliation{Laboratoire de Photonique et de Nanostructures (LPN), CNRS, Université Paris-Saclay, route de Nozay, F-91460 Marcoussis, France}

\author{I. Sagnes}
\affiliation{Laboratoire de Photonique et de Nanostructures (LPN), CNRS, Université Paris-Saclay, route de Nozay, F-91460 Marcoussis, France}

\author{S. Schmidt}
\affiliation{Institute for Theoretical Physics, ETH Zurich, 8093 Zurich, Switzerland}

\author{H. E. T\"ureci}
\affiliation{Department of Electrical Engineering, Princeton University, Princeton, New Jersey 08544, USA}

\author{A. Amo}
\affiliation{Laboratoire de Photonique et de Nanostructures (LPN), CNRS, Université Paris-Saclay, route de Nozay, F-91460 Marcoussis, France}

\author{J. Bloch}
\affiliation{Laboratoire de Photonique et de Nanostructures (LPN), CNRS, Université Paris-Saclay, route de Nozay, F-91460 Marcoussis, France}
\affiliation{Physics Deparment, Ecole Polytechnique, Université Paris-Saclay, F-91128 Palaiseau Cedex, France}

\begin{abstract}

We report on the engineering of a non-dispersive (flat) energy band in a geometrically frustrated lattice of micro-pillar optical cavities.
By taking advantage of the non-hermitian nature of our system, we achieve bosonic condensation of exciton-polaritons into the flat band. 
Due to the infinite effective mass in such band, the condensate is highly sensitive to disorder and fragments into localized modes reflecting the elementary eigenstates produced by geometric frustration.
{\FB This realization} offers a novel approach to studying coherent phases of light and matter under the controlled interplay of frustration, interactions and dissipation.

\end{abstract}

\maketitle

Flat energy bands occur in a variety of condensed matter systems, from the Landau levels of an electron gas, edge states of graphene \cite{Nakada96}, Aharonov-Bohm cages in metal networks \cite{Naud01}, frustrated magnets \cite{Ramirez94} to unconventional superconductors \cite{schnyder11}. The common feature of these materials is the appearance of a divergence in the density of states at the energy of the flat band, which prevents straightforward ordering.  Consequently, any small perturbation may have a dramatic effect on the system. For example, interactions often lead to strongly correlated and exotic phases of matter as observed in the fractional quantum Hall effect~\cite{DePicciotto97}, spin liquids \cite{Lee02} or spin ices \cite{ramirez99,Isakov05}. Another important class of emerging phenomena in flat band systems originates from the effects of disorder, which are enhanced by the very large mass \cite{Faggiani15} and can significantly deviate from conventional Anderson localization. Examples include the inverse Anderson transition (delocalization transition) \cite{Goda06}, localization with unconventional critical exponents and multi-fractal behavior \cite{chalker10}, and mobility edges with algebraic singularities~\cite{Bodyfelt14}.

The observation of these phenomena in solid-state systems is often complicated by extrinsic material-specific perturbations and the impracticality of engineering suitable lattice geometries. Artificial lattices, recently implemented in a number of physical systems, allow simulating this flat band physics in a controllable manner. Pioneering works in photonic systems  \cite{Nakata12,Masumoto12,Nixon13,Jacqmin14,GuzmanSilva14,Vicencio15,Mukherjee15}
and cold atom gases \cite{Jo12,Aidelsburger15,Taie15} highlighted the key role played by geometric frustration  \cite{Nixon13} and evidenced characteristic features such as the absence of {\FB wavepacket diffraction} in a flat band~\cite{GuzmanSilva14,Vicencio15,Mukherjee15}.
But {\FB despite many predictions \cite{Vicencio13,Molina12,Leykam13,Ge15} } the specific role played by interactions or disorder has not been {\FB experimentally} addressed in these artificial lattices so far.
Also, steady-state bosonic condensation has not been realized in a flat band, and the question of the experimental properties of such condensate remains open. In the case of ultracold atomic systems such study remains delicate as flat bands usually appear at energies far above the ground state \cite{huber10,Jo12,Tovmasyan13}.
Investigation of flat band condensation thus require challenging experimental techniques to implement complex-valued tunneling constants \cite{Aidelsburger15} or the coherent transfer of the atomic condensate into an excited state \cite{Taie15}.

\begin{figure}[h!]
\includegraphics[width=0.95\columnwidth]{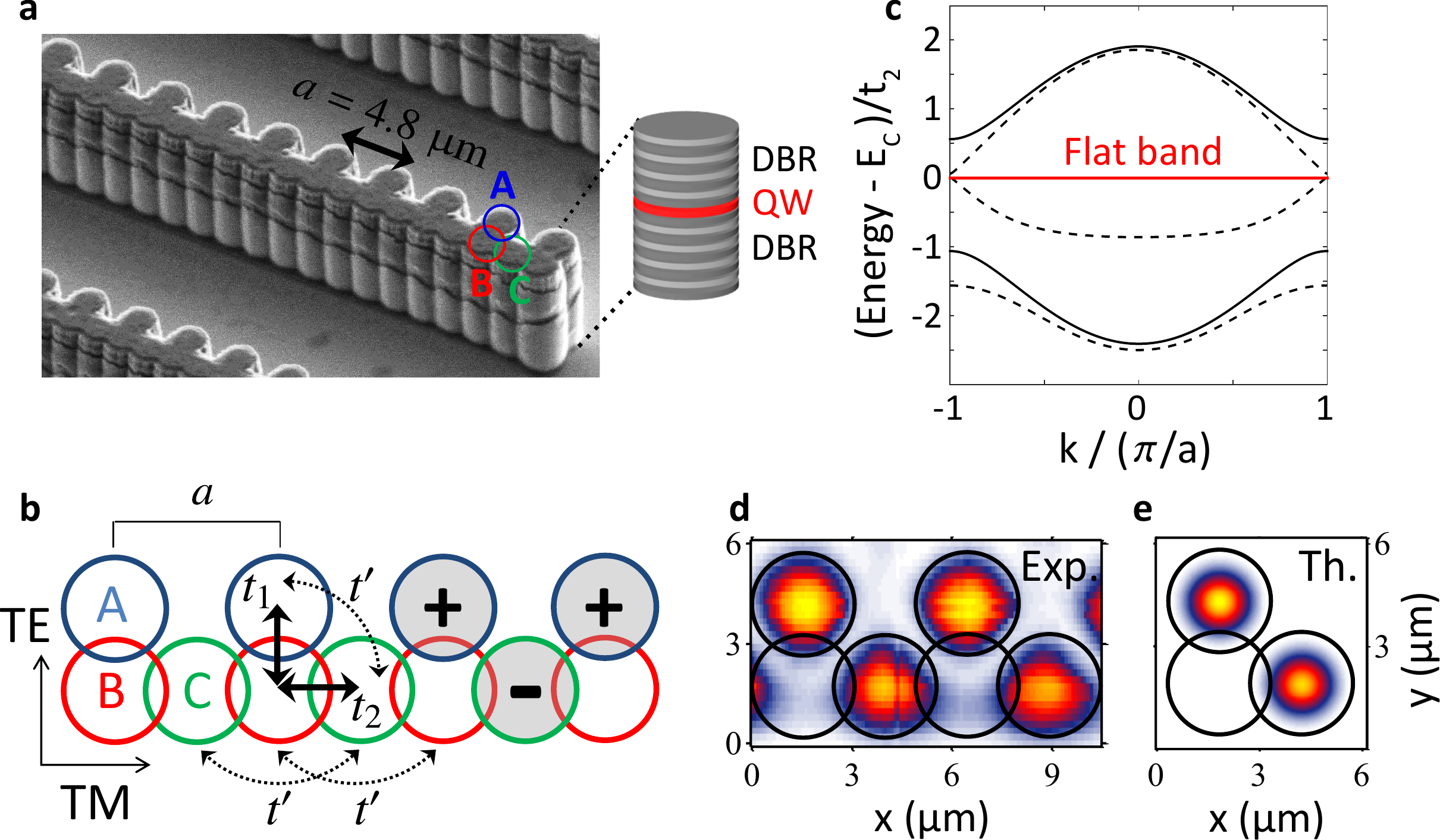}
\caption{
\textbf{a}, Scanning electron micrograph of a 1D Lieb {\FB (or Stub)} lattice of coupled micropillars etched out of a cavity. The close up shows schematically the pillars structure, with two distributed Bragg reflectors (DBR) forming the cavity, and quantum wells (QW) inserted at the antinodes of the optical field.
\textbf{b}, Geometric structure of the lattice: it contains three types of pillars (A,B,C) linked by couplings $t_1,t_2$ to nearest neighbors  and $t'$ to next-nearest neighbors. An example of plaquette-state (see text) is highlighted in grey.
\textbf{c}, Band structure calculated in the tight-binding model with $t'=0$, at zero and finite spectral detuning between A and C sites (solid and dotted lines, respectively). 
\textbf{d}, Experimental real space emission at the flat band energy (conditions of Fig. \ref{Fig2}b). 
\textbf{e}, Calculated density profile of a typical flat band mode (only one unit cell is shown).
}
\label{Fig1}
\end{figure}

In the present work, we use exciton-polaritons to investigate bosonic condensation in a flat band.
These quasiparticles arise from the strong coupling between excitons confined in quantum wells and photons confined in a semiconductor microcavity \cite{Weisbuch92}. Their mixed light-matter nature allows efficient band structure engineering through their photonic component \cite{Bayer99}, while providing scattering channels and non-linearities through their excitonic component \cite{Tassone99}. 
These assets have allowed studying polaritons in staggered \cite{CerdaMendez10,Tanese13}, squared \cite{Kim11}, honeycomb \cite{Jacqmin14} and Kagome lattices \cite{Masumoto12}.
In addition, their dissipative nature makes them an archetypal non-hermitian system \cite{Ge13,Gao15,MoiseyevBook} featuring novel dynamical universality classes \cite{Sieberer13}.
In this Letter, we engineer geometric frustration by patterning a cavity into a 1D Lieb lattice {\FB (also known as Stub lattice)} \cite{Hyrkas13} of optical micro-pillars. Photoluminescence at low excitation power directly reveals the presence of a flat energy band. 
At higher excitation power, we demonstrate the ability to trigger bosonic condensation of polaritons in the flat band. Combined interferometric and spectral measurements reveal that condensation occurs on highly localized modes, reflecting the elementary eigenstates produced by geometric frustration. These localized modes arise from disorder in the sample, to which flat band states are extremely sensitive due to their infinite effective mass.

Our 1D Lieb lattice of coupled micro-pillars [Fig.~\ref{Fig1}a] is obtained by processing a planar microcavity  {\FB (of nominal Q-factor 70000)} grown by molecular beam epitaxy. The cavity consists of a $\lambda / 2$ Ga$_{0.05}$Al$_{0.95}$As layer surrounded by two Ga$_{0.2}$Al$_{0.8}$As/Ga$_{0.05}$Al$_{0.95}$As Bragg mirrors with 28 and 40 pairs in the top/bottom mirrors, respectively. Twelve GaAs quantum wells of width 7 nm are inserted in the structure, resulting in a 15 meV Rabi splitting.
Micropillars (see close-up in Fig.~\ref{Fig1}a) are patterned by dry etching down to the GaAs substrate. The diameter of each pillar is $3$ $\mu$m, and
the distance between two adjacent pillars is $2.4$ $\mu$m, so that they spatially overlap, allowing for the tunneling of polaritons \cite{Bayer99,Jacqmin14}. 

\begin{figure}[h!]
\includegraphics[width=0.95\columnwidth]{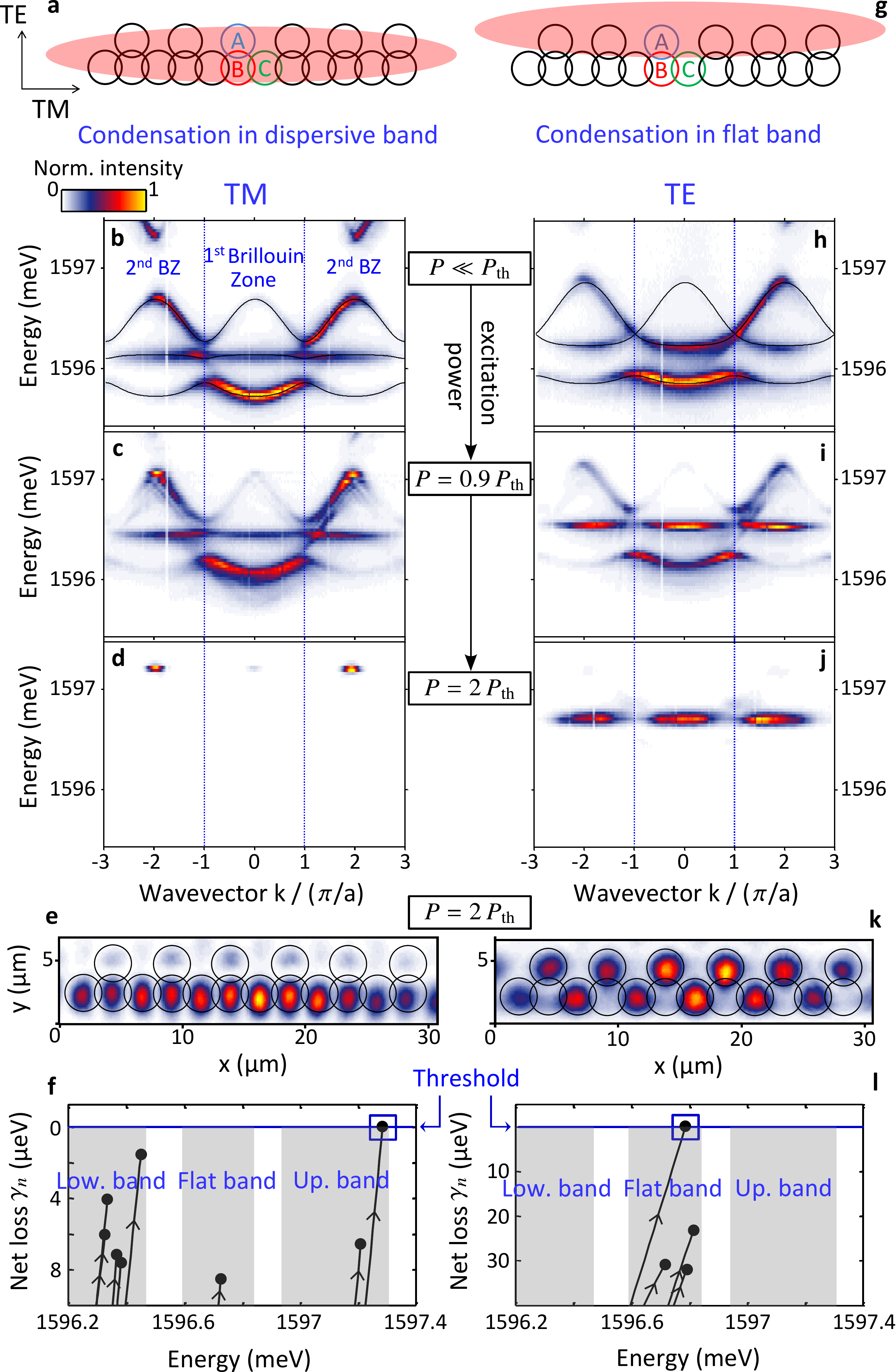}
\caption{
\textbf{b,h}, Energy-resolved far field emission of the structure at low excitation power, for polarization parallel (TM) and perpendicular (TE) to the lattice. Black lines are fits using the tight-binding Hamiltonian $\mathcal{H}_{\mathrm{Lieb}}$ [see Eq. \eqref{Eq1}]. 
When the pump excites equally all type of pillars (\textbf{a}), condensation occurs in the upper dispersive band and is TM polarized: \textbf{b}-\textbf{d} monitor the far field emission when increasing the pump power from $P=0.05 \, P_{\rm th}$ to $P=2 \, P_{\rm th}$, where $P_{\rm th}$ is the threshold power ($P_{\rm th} \simeq 4$ mW for this condensation process). In \textbf{d} condensation is achieved at the top of the upper dispersive band.
\textbf{e}, Corresponding real space image of the condensate.
\textbf{f},~Calculated trajectory of the eigenvalues of Eq. \eqref{Eq1} in the complex plane, as the power is increased (see arrows). 
\textbf{g},~When the pump is centered on the line of pillars A, condensation occurs in the flat band and is TE polarized: \textbf{h}-\textbf{j} monitor the far field emission with increasing pump power ($P_{\rm th} \simeq 5$ mW for this condensation process).
\textbf{k},~Corresponding real space image of the flat band condensate.
\textbf{l},~Same calculation as \textbf{f} but for the pump configuration~\textbf{g}, reproducing condensation in the flat band.}
\label{Fig2}
\end{figure}

\begin{figure*}[t]
\includegraphics[width=1\textwidth]{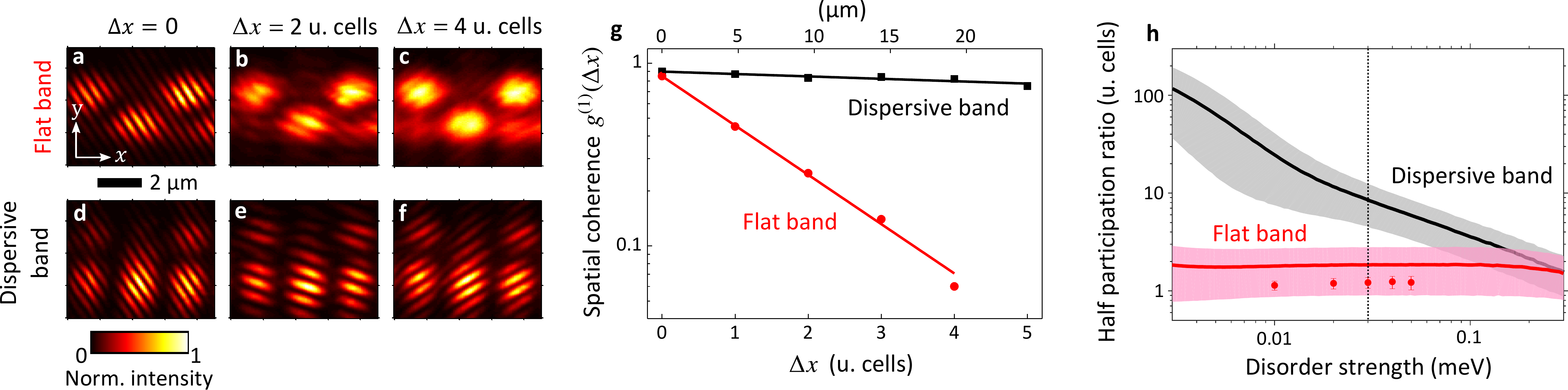}
\caption{
 \textbf{a-c}, Interferograms obtained by superposing two images of the flat band condensate (at $P=2 \,P_{\rm th}$) shifted by $\Delta x=0$, $2$ and $4$ unit cells. \textbf{d-f},~Same experiment realized with the upper dispersive band condensate.
\textbf{g}, First-order spatial coherence  $g^{(1)}$ deduced from the visibility of the interferograms, as a function of $\Delta x$. 
\textbf{h}, Solid lines: Half participation ratio $\HPR$ calculated from $\mathcal{H}_{\mathrm{Lieb}}$ as a function of (diagonal) disorder strength, in the flat band and at the top ($k=0$) of the upper dispersive band. Shaded aeras  denote the standard deviation, and the vertical line indicates the estimated disorder strength in our sample. Dots: $\HPR$ in the flat band calculated from the full Hamiltonian $\mathcal{H}(P)$ [Eq. \eqref{Eq1}].
}
\label{Fig3}
\end{figure*}

The unit cell of the 1D Lieb lattice (Fig. \ref{Fig1}b) contains three sites (A,B,C) linked by a coupling constant $t_1$ between A and B, and $t_2$ between B and C. 
In general, this lattice exhibits three dispersive bands, as shown in the tight-binding calculation of Fig. \ref{Fig1}c (dashed lines). 
In the particular case of sites A and C having equal energies, the middle band is flat and gapped from the two remaining dispersive bands (solid lines). 
The existence of the flat band is independent of the energy of pillars B and of the couplings $t_1,t_2$.

To probe the polariton properties in this lattice, we excite it non-resonantly with a continuous-wave monomode laser at 740 nm. The spot is elliptical with $2$ $\mu$m-width and $60$ $\mu$m-length (intensity FWHM), corresponding to $12$ unit cells of the lattice. Experiments are performed at 6K and $-8$ meV cavity-exciton detuning. 
The emission of the sample is collected with a $0.5$ numerical aperture objective and focused on the entrance slit of a spectrometer coupled to a CCD camera. Imaging of the sample surface (resp. the Fourier plane) allows for studying the polariton modes in real (resp. reciprocal) space.  
We resolve the emission in polarization, and consider either the polarization
parallel (TM) or perpendicular (TE) to the lattice (Fig. \ref{Fig1}b).

Under low pumping power, incoherent relaxation of polaritons results in the population of all energy bands, allowing a direct visualization of the band structure. {\FB The corresponding far field emission is shown in Figs. \ref{Fig2}b and \ref{Fig2}h for emitted light polarized TM and TE, respectively.} 
In TM polarization a gapped flat band is clearly observed, while in TE polarization the middle band is dispersive and crosses the upper band. 
This difference arises from polarization-dependent boundary conditions for the photonic modes of the pillars, which induce a spectral detuning between pillars A and C for TE polarization.
{\FB Both band structures are~well reproduced (black lines in Figs.~\ref{Fig2}b,h) by a tight-binding~Hamiltonian  including next-nearest-neighbor couplings~$t^{\prime}$: $\mathcal{H}_{\mathrm{Lieb}}=$  
 $\Scale[0.92]{\sum_{l,j}E_{l}\left\vert l_j\right\rangle \left\langle l_j\right\vert
 -\sum_{j}(t_{1}\left\vert A_j\right\rangle \left\langle B_j\right\vert
 +t_{2}\left\vert B_j\right\rangle \left\langle C_j\right\vert+t_{2}\left\vert C_j\right\rangle \left\langle B_{j+1} \right\vert )}$
 $\Scale[0.92]{-\sum_{j}t^{\prime }(\left\vert B_j\right\rangle  \left\langle
B_{j+1}\right\vert + \left\vert C_j\right\rangle  \left\langle
C_{j+1}\right\vert 
+ \left\vert  A_{j}\right\rangle \left\langle
 C_{j}\right\vert
+ \left\vert  C_{j}\right\rangle \left\langle
 A_{j+1}\right\vert) \!+h.c.}\,$,
 where $\left\vert l_j\right\rangle $ denotes the wavefunction (of energy $E_{l}$) of pillar $l=A,B,C$ in the $j^{\rm th}$ unit cell.}

{\FB The real space pattern corresponding to the flat band of Fig. \ref{Fig2}b (i.e. below the condensation threshold) can be reconstructed by spectrally filtering the image. The result is shown in Fig. \ref{Fig1}d, and compared to the tight-binding prediction of Fig. \ref{Fig1}e. A characteristic structure is observed, with sites B being dark, i.e., containing no particles. This is a direct signature of geometric frustration: a destructive interference between sites A and C cancels the net flow of particles to sites B and prevents propagation of flat band states \cite{huber10,Hyrkas13}.}

We now consider polariton condensation \cite{Kasprzak06} in this 1D Lieb lattice. Upon increasing  the pump power, condensation is triggered when the relaxation towards a given mode becomes faster than its decay \cite{Imamgoglu96}. 
The mode becomes macroscopically occupied and spontaneous coherence sets in. 
This process can be understood by analyzing the non-Hermitian operator derived from the generalized Gross-Pitaevskii equation \cite{Carusotto13}:
{\FB
\begin{multline}
\Scale[0.94]{\mathcal{H}(P)=\mathcal{H}_{\mathrm{Lieb}}}
\Scale[0.94]{+\!\!\!\!\!\!\sum\limits
_{\substack{ l=A,B,C \\ j=1...N}}\left[ \frac{g_{R}}{\gamma _{R}}f_{lj}P+%
\frac{i}{2}\left( \frac{R}{\gamma _{R}}f_{lj}P-\gamma _{l}\right)
\right] \left\vert l_j\right\rangle \left\langle l_j\right\vert} 
\label{Eq1}
\end{multline}
The second part in Eq. \eqref{Eq1} is
complex-valued and describes the effect of pump and dissipation which adds to the lattice Hamiltonian $\mathcal{H}_{\mathrm{Lieb}}$ that is fixed by the fabricated structure. The first
term accounts for the repulsive interaction between polaritons and reservoir
excitons, with $g_{R}$ being the corresponding interaction constant, $\gamma
_{R}$ and $R$ the exciton decay and relaxation rates, respectively. This repulsive interaction produces a
blueshift of the pillar energies, proportional to the pump power $P$;  $f_{lj}
$ denotes the fraction of the total pump power on each pillar. The second
term describes the dissipation rate of each pillar: it is the sum of the
passive dissipation rate $\gamma _{l}$ imposed by polariton decay, and a
gain term proportional to $P$.}

The eigenvalues of $\mathcal{H}(P)$ are complex-valued $\omega_n=\nu_n - i \gamma_n /2$ and describe the frequency and damping of the linear fluctuations around the uncondensed state \cite{Ge13}.
As the power $P$ is ramped up, all complex eigenvalues flow towards the real axis from below (see Figs. \ref{Fig2}f,l). The condensation threshold for a given mode is reached when its eigenvalue $\omega_n$ crosses the real axis (net loss $\gamma_n=0$), meaning that gain overcomes the decay of polaritons.
Whereas the polariton decay is fixed by the parameters of our structure, the gain is proportional to the spatial overlap with the pump and can thus be controlled by tailoring the pump profile {\FB(described by $f_{lj}$)} \cite{Ge13}. This is indeed what we demonstrate in Fig. \ref{Fig2}: in the 1D Lieb lattice we can control condensation to occur either in the upper dispersive band (left column) or in the flat band (right column), by adjusting the pump spatial configuration.

We first pump the lattice in a symmetric manner, as sketched in Fig. \ref{Fig2}a: all types of micropillars (A,B,C) are pumped by the same amount ($f_{Aj}\!\!=\!\! f_{Bj}\!\!=\!\! f_{Cj}$). 
Figs.~\ref{Fig2}b-d monitor the momentum space emission in TM polarization, when increasing the pump power from $P=0.05\, P_{\rm th}$ to $P=2\, P_{\rm th}$, where $P_{\rm th}= 4$ mW is the threshold power (corresponding data for TE polarization are shown in the Supplemental Material \cite{SM}).
 Polariton condensation eventually occurs in the upper dispersive band (Fig. \ref{Fig2}d), with the emission collapsing at the center of the Brillouin zones \cite{Jacqmin14} and showing narrow linewidth.
This condensation process is well reproduced by calculating the trajectory of the eigenvalues of $\mathcal{H}(P)$ in the complex plane (see Fig. \ref{Fig2}f). 
{\FB The parameters of $\mathcal{H}_{\rm Lieb}$ are kept constant and the evolution of the eigenvalues with pump power is completely driven by the second part of Eq. \eqref{Eq1}.
We observe that the $k=0$ upper band mode (enclosed by the square box) indeed reaches the real axis first for this pump configuration.}

In the following, we show that asymmetric pumping of the same structure allows triggering condensation into the flat band. Compared to upper band modes, flat band modes have larger amplitude on pillars A (Fig.~\ref{Fig1}d,e). Pumping favorably these pillars ($f_{Aj} \!\gg \!f_{Bj}, f_{Cj}$) thus enhances the gain and lowers the condensation threshold of flat band modes, as confirmed by the calculated spectrum in Fig. \ref{Fig2}l.
Using such asymmetric pumping, pillars A develop an additional blueshift with respect to pillars C, which destroys the flat band in TM polarization \cite{SM}.
Nevertheless, in TE polarization this blueshift compensates the photonic detuning between pillars A and C so that a gapped flat band is formed, as experimentally observed in Fig.~\ref{Fig2}i. Increasing the power further triggers condensation in this flat band (Fig.~\ref{Fig2}j). The emission in real space (Fig.~\ref{Fig2}k) shows dark B sites characteristic of geometric frustration, in clear contrast with the dispersive band condensate (Fig.~\ref{Fig2}e).

A peculiarity of a flat band is that, due to its macroscopic degeneracy, eigenstates of arbitrary lengths can be formed. 
Among these, a set of maximally localized eigenmodes can be defined \cite{huber10}, which for the Lieb lattice extend over 3 pillars only \cite{Hyrkas13}.
 An example of such a \textit{plaquette-state} is highlighted in gray in Fig. \ref{Fig1}b.
Note that delocalized eigenstates can also be obtained by linear superposition of different plaquette-states.
Thus, the question arises of which kind of state is actually picked up for condensation in the flat band.

\begin{figure}[h!]
\includegraphics[width=0.8\columnwidth]{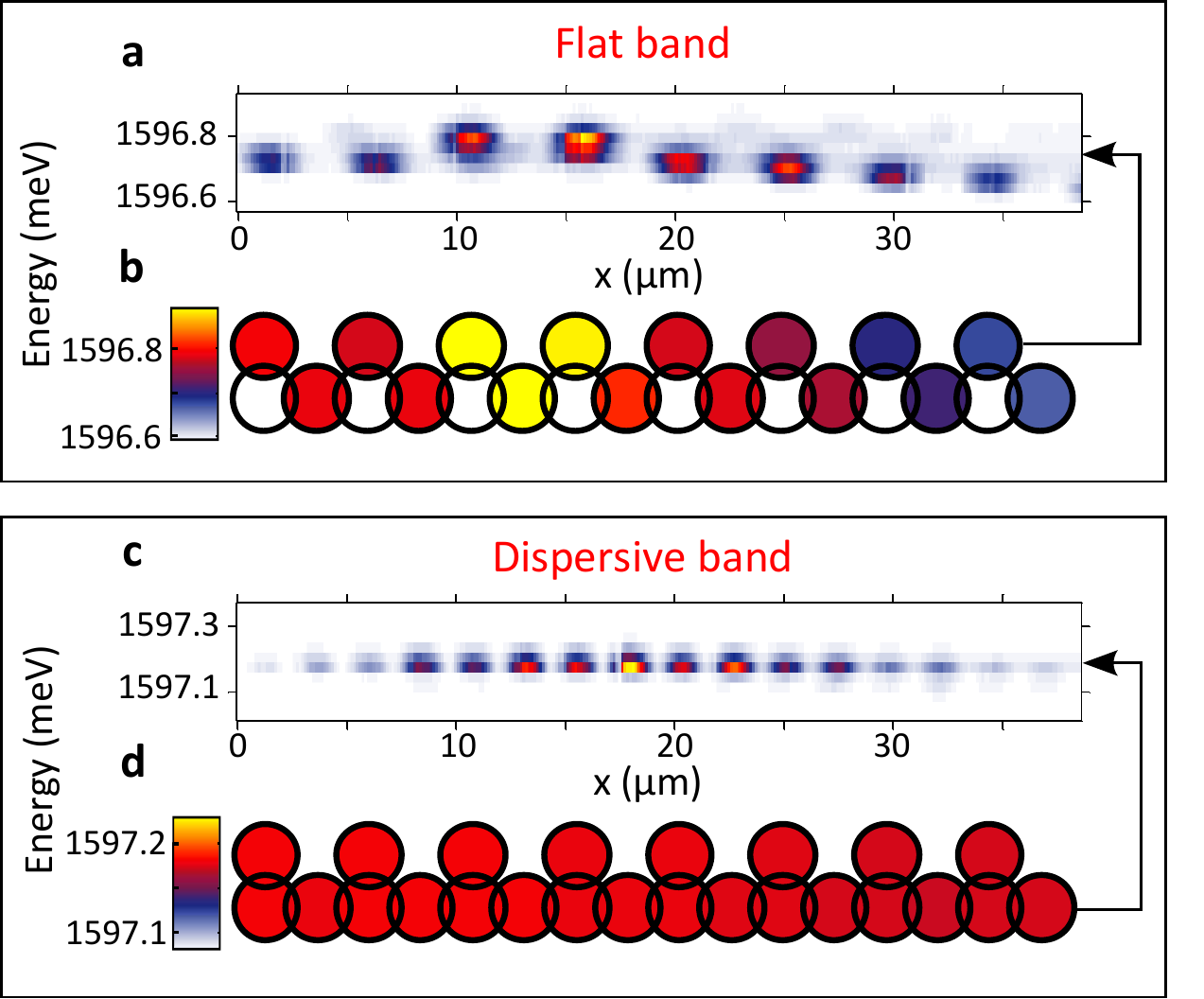}
\caption{
\textbf{a}, Energy-resolved emission of the flat band condensate in real space (see Fig. \ref{Fig2}k) along the line of pillars A.
We colour-code in \textbf{b} the emission energy along the lattice: several localized modes are visible, emitting at distinct energies. 
\textbf{c}, Energy-resolved emission of the upper dispersive band condensate in real space (see Fig. \ref{Fig2}e), along the line of pillars B and C. 
\textbf{d}, Corresponding spatial map of the emission energy, showing monochromatic emission over the whole condensate.
}
\label{Fig4}
\end{figure}

To answer this question, we use interferometric measurements and high spectral resolution imaging. We first investigate the first-order spatial coherence $g^{(1)} (\Delta x)$ of the flat band emission, by superposing two images shifted by $\Delta x$ along the periodic direction, and extracting the averaged visibility of the resulting interference fringes \cite{Kasprzak06}. Figs.~\ref{Fig3}a-c show the measured interferograms for increasing values of $\Delta x$: the visibility decays exponentially (Fig.~\ref{Fig3}g) with a characteristic length of $1.6 \pm 0.3$ unit cells.  This indicates the presence of many independent condensates, localized on 1.6 unit cells on average, i.e. close to plaquette-states. To confirm this picture we use high spectral resolution imaging. Fig.~\ref{Fig4}a displays the energy-resolved emission of the flat band condensate in real space, along the line of pillars A. By repeating the measurement for the line of pillars B and C \cite{SM}, we color-code in Fig.~\ref{Fig4}b the emission energy measured on each pillar. It shows significant spectral variations along the lattice, allowing to visualize the localized condensates. 
{\FB A general measure of the localization of these condensates is provided by the half participation ratio, defined as $\HPR=(\sum_j |\psi_{Cj}|^2)^2 / (2 \sum_j |\psi_{Cj}|^4)$, where $\psi_{Cj}$ is the wavefunction amplitude on site $C$ of the $j^{\rm th}$ unit cell. The $\HPR$ quantifies the portion of the lattice where the wavefunction differs markedly from zero, and coincides with the localization length in the case of exponentially localized states~\cite{kramer1993}.}
In Fig. \ref{Fig4}b, starting from the left, one successively identifies flat band condensates with an $\HPR$ of $\sim 1.5$, $1$ and $2$ unit cells. This confirms that flat band condensation is multimode, occurring simultaneously on strongly localized and independent modes.

Let us compare these observations to the situation where condensation takes place in the upper dispersive band. Keeping the very same location on the sample, Figs.~\ref{Fig3}d-f show the corresponding interferograms obtained when the system is pumped symmetrically (Fig. \ref{Fig2}a). 
The deduced $g^{(1)}$ decays much more slowly (Fig. \ref{Fig3}g), revealing highly extended spatial coherence.
Consistently, the energy-resolved measurements of Fig. \ref{Fig4}c-d show that the emission is monochromatic over the whole condensate within the experimental resolution (30~$\mu$eV): condensation here occurs on a single extended mode, having an $\HPR>4$ unit cells.

This striking difference between both types of condensates can be understood by taking system disorder into account.
{\FB A flat band is robust to non-diagonal disorder (i.e. on the tunnel couplings) \cite{Mukherjee15}, but any
small amount of diagonal disorder (on the on-site energies) breaks the macroscopic degeneracy: extended eigenstates cannot be formed anymore, resulting in a strong localization effect \cite{chalker10,Leykam13,Faggiani15}. 
This can be seen by calculating the eigenstates of the lattice $\mathcal{H}_{\rm Lieb}$ in presence of diagonal disorder.
We plot in Fig. \ref{Fig3}h the corresponding half participation ratio $\HPR$.
In the flat band (red line) the $\HPR$ is very small and essentially independent of the disorder strength, illustrating the non-perturbative effect of disorder. Its value $\HPR \simeq 1.8 \pm 1$ unit cells reproduces well the experimental $\HPR$.
By contrast, the $\HPR$ in the upper dispersive band (black line) decays as a power law with increasing disorder, and is several times larger than in the flat band for weak disorder. We calculate $\HPR = 8.5 \pm 4$ for the estimated disorder strength of our sample (30~$\mu$eV, vertical line in Fig. \ref{Fig3}h), which is compatible with the extended condensate experimentally observed.

{\FB To investigate the possible influence of pump and dissipation on the strong localization observed in the flat band, we also calculate the $\HPR$ using the complete Hamiltonian $\mathcal{H}(P)$ [Eq. \eqref{Eq1}]. 
In absence of disorder, we find that the $\HPR$ in the flat band is bounded only by the finite size of the pump spot~\footnote{The $\HPR$ tends to the lattice size when pumping all pillars.} (as is the case for the upper band), showing that pump and dissipation alone cannot lead to the observed localization.
When including disorder, the $\HPR$ (red dots in Fig. \ref{Fig3}h) is close to the calculation based on $\mathcal{H}_{\rm Lieb}$ only.} 
This allows concluding that pump and dissipation play a minor role in the observed localization effect. We note that polariton-polariton interactions (cubic nonlinearity) could also lead to localization in the flat band \cite{Vicencio13,Molina12,Leykam13}, but in the current experiment this nonlinearity is of order $\sim 0.1$ $\mu$eV and thus much smaller than disorder (30 $\mu$eV) \cite{SM}. {\FB We thus neglected this term in the analysis.}

Finally, the experimental reproducibility of our results has been verified on ten different lattice realizations. Each time the disorder landscape, which allows the formation of an extended monomode condensate in the upper band, leads to a fragmentation of the flat band condensate into highly localized modes. To conclude, our experiments demonstrate the extreme sensitivity of a flat band condensate to diagonal disorder, due to the infinite effective mass in such band.}
Bosonic condensation in a flat energy band opens up novel perspectives in the simulation of mesoscopic and condensed matter phenomena.
In particular, access to many-body flat band physics \cite{Wu07,huber10,biondi15} could be gained using resonant pumping, where the interaction energy is set by the spectral detuning between the pump and the considered quantum state \cite{Baas04}.
Furthermore, spin-orbit coupling has recently been engineered in polaritonic systems \cite{Sala15}. This should allow to further explore the delicate interplay of frustration, interactions and topology \cite{Sun11,Liu12} in macroscopic quantum states.

\bibliography{Biblio}

\bigskip
\bigskip
\textbf{Acknowledgments}: We thank Paul Voisin for fruitful discussions. This work was supported by
the Agence Nationale de la Recherche project \textit{Quandyde} (Grant No. ANR-11-BS10-001), the French RENATECH network, the European Research Council grant \textit{Honeypol}, the EU-FET Proactive grant \textit{AQuS} (Project No. 640800), the NSF Grant No. DMR-1151810, the Swiss NSF through an Ambizione Fellowship under Grant No. PZ00P2\_142539, the CUNY grant CIRG-802091621 and the NCCR QSIT.

\end{document}